\documentclass[11pt,english]{article}
\usepackage[]{graphicx}
\usepackage{graphics, epsfig}
\usepackage{amssymb}
\usepackage{amsmath}

\title{A Snapshot of J. L. Synge}

\author{Peter A. Hogan\footnote{Email: peter.hogan@ucd.ie}, \\
        School of Physics\\
        University College Dublin\\
        Belfield, Dublin 4, Ireland}
        
\date{}

\begin{document}
\maketitle

\begin{figure}
\centering
\includegraphics[scale=.6]{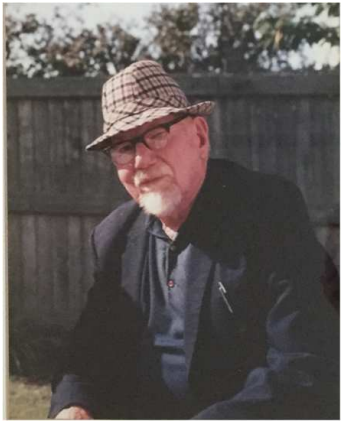}
\caption{J. L. Synge 1897-1995}
\label{Synge_picture}
\end{figure}

\begin{abstract}
A brief description is given of the life and influence on relativity theory of Professor J. L. Synge accompanied by some technical examples to illustrate his style of work.
\end{abstract}

\section{Introduction}
When I was a postdoctoral fellow working with Professor Synge in the School of Theoretical Physics of the Dublin Institute for Advanced Studies he was fifty--one years older than me 
and he remained research active for another twenty years. John Lighton Synge FRS was born in Dublin on 23rd. March, 1897 and died in Dublin on 30th. March, 1995. As well as his emphasis on, 
and mastery of, the geometry of space--time he had a unique delivery, both verbal and written, which I will try to convey in the course of this short article. But first the basic facts of his academic life are 
as follows: He was educated in St. Andrew's College, Dublin and entered Trinity College, University of Dublin in 1915. He graduated B.A. (1919), M.A. (1922) and Sc.D. (1926). He was Assistant Professor 
of Mathematics in the University of Toronto (1920--25), subsequently returning to Trinity College Dublin as Professor of Natural Philosophy (1925--30) and then left for the University of Toronto again to take up the position of 
Professor of Applied Mathematics (1930--43). From there he went to Ohio State University as chairman of the Mathematics Department (1943--46) followed by Head, Mathematics Department at Carnegie 
Institute of Technology, Pittsburgh (1946--48) before returning to Dublin to establish his school of relativity in the Dublin Institute for Advanced Studies. He officially retired when he was seventy--five years old. 

Synge was prolific, publishing 250 papers and 11 books. In 1986 he wrote, but did not publish, some informal autobiographical notes \cite{Synge/AutobioNotes}, which he described as being for his family and descendants 
and to aid obituary writers, and which are deposited in the library of the School of Theoretical Physics of the Dublin Institute for Advanced Studies. 

Early in his career he published his first important paper: ``On the Geometry of Dynamics", {\it Phil. Trans. Roy. Soc.} A{\bf 226} (1926), 31-106. Of this work he said 
\cite{Synge/AutobioNotes}; ``I sent a copy to T. Levi-Civit\`a, and in return he sent me a copy of a paper by him, just appearing. Our papers had in common the equation 
of geodesic deviation, now familiar to relativists, but he had done it using an indefinite line element, appropriate to relativity, whereas my line element was positive definite." 

\setcounter{equation}{0}
\section{A Scheme of Approximation}\indent 
Synge placed great emphasis on working things out for oneself, writing that \cite{RGT} ``the lust for calculation must be tempered by periods of inaction, in which the mechanism is completely unscrewed and then put together again. It is the decarbonisation of the mind." As an illustration of this activity I give a weak field approximation scheme published in 1970 by Synge \cite{Approx} which has the advantage that it can be described without reference to an example. This is a topic which, by 1970, had become a standard entry in textbooks on general relativity and one might be forgiven for thinking that by then the last 
word had been said on it. 

We first need some basic objects and notation. In Minkowskian space-time Synge liked to use imaginary time (which some people find maddening!) and to write the position 4--vector in rectangular Cartesians and time as
\begin{eqnarray}
x_a=(x, y, z, it)\ \ {\rm with}\ \ a=1, 2, 3, 4\ \ {\rm and}\ \ \ i=\sqrt{-1},\nonumber\end{eqnarray}with the index in the covariant or lower position. The Minkowskian metric tensor in these coordinates has components $\delta_{ab}$ (the Kronecker delta). If the metric tensor of a space--time has components of the form
\begin{eqnarray}
g_{ab}=\delta_{ab}+\gamma_{ab}\ ,\nonumber\end{eqnarray}then he defined the ``truncated Einstein tensor" $\hat G^{ab}$ via
\begin{eqnarray}
G^{ab}=L_{ab}+\hat G^{ab}\ ,\nonumber\end{eqnarray}where $G^{ab}$ is the Einstein tensor calculated with the metric $g_{ab}$ and 
\begin{eqnarray}
L_{ab}=\frac{1}{2}(\gamma_{ab,cc}+\gamma_{cc,ab}-\gamma_{ac,cb}-\gamma_{bc,ca})-\frac{1}{2}\delta_{ab}(\gamma_{cc,dd}-\gamma_{cd,cd})\ .\nonumber\end{eqnarray}
The energy--momentum--stress tensor of matter giving rise to a gravitational field has components $T^{ab}$. With these preliminaries Synge's strategy is as follows: 
\vskip 1truepc\noindent
(1) Given $T^{ab}$, generate a sequence of metrics $\underset{M}{g_{ab}}=\delta_{ab}+\underset{M}{\gamma_{ab}}\ \ \ (M=0, 1, 2, \dots\ \ , N)$;
\vskip 1truepc\noindent
(2) Approximations are introduced by expressing the components $T^{ab}$ in terms of a small parameter;
\vskip 1truepc\noindent
(3) Integrability conditions, equivalent to the equations of motion, are imposed to terminate the sequence at a term which satisfies Einstein's field equations with a predicted order of approximation in terms of the small parameter.
\vskip 1truepc\noindent
The sequence is constructed as follows: with
\begin{eqnarray}
\underset{M}{\gamma^*_{ab}}=\underset{M}{\gamma_{ab}}-\frac{1}{2}\delta_{ab}\,\underset{M}{\gamma_{cc}}\ \ \ (M=0, 1, 2, \dots\ \ N)\ ,\nonumber\end{eqnarray}
and
\begin{eqnarray}
\underset{M}{H}^{ab}=T^{ab}+(8\,\pi)^{-1}\underset{M}{\hat G}^{ab}(\underset{M}{\gamma} )\ \ \ (M=0, 1, 2, \dots\ \ N)\ ,\nonumber\end{eqnarray}
\emph{define the sequence} $\{\underset{M}{\gamma_{ab}}\}$ by
\begin{eqnarray}
\underset{0}{\gamma_{ab}}=0\ \ \ {\rm and}\ \ \ \underset{M}{\gamma^*_{ab}}=16\,\pi\,K^{ab}_{rs}\underset{M-1}{H^{rs}}\ \ \ (M=1, 2, 3, \dots\ \ N)\ .\nonumber\end{eqnarray}Here $K^{ab}_{rs}$ is an operator defined by
\begin{eqnarray}
K^{ab}_{rs}=-\delta_{ar}\,\delta_{bs}\,J+J(\delta_{ar}\,D_{bs}+\delta_{bs}\,D_{ar}-\delta_{ab}\,D_{rs})J\ ,\nonumber\end{eqnarray}
with $D_a=\partial/\partial x_a$, $D_{ab}=\partial^2/\partial x_a\partial x_b$. The operator $J$ is the inverse d'Alembertian:
\begin{eqnarray}
Jf(\vec x, t)=-\frac{1}{4\,\pi}\int\frac{f(\vec x', t-|\vec x-\vec x'|)}{|\vec x-\vec x'|}d_3x'\ .\nonumber\end{eqnarray}
Synge proved that the integrals involved in the implementation of the operator $K$ converge if the physical system is stationary ($T^{ab}{}_{,4}=0$) for some period in the past. He called this property $J$--convergence.
\vskip 1truepc\noindent
Approximations are introduced as follows: all $\{\underset{M}{\gamma_{ab}}\}$ defined above satisfy the coordinate conditions
\begin{eqnarray}
\underset{M}{\gamma^*_{ab,b}}=0\ \ (M=0, 1, 2, \dots\ \ N)\ .\nonumber\end{eqnarray}
Introduce approximations by assuming $T^{ab}=O(k)$ for some dimensionless parameter $k$ then
\begin{eqnarray}
\underset{M}{\gamma_{ab}}-\underset{M-1}{\gamma_{ab}}=O(k^M)\ \ \ (M=1, 2, 3, \dots\ \ ,N)\ ,\nonumber\end{eqnarray}
from the definition of $\gamma_{ab}$, and 
\begin{eqnarray}
\underset{M}{\hat G}^{ab}-\underset{M-1}{\hat G}^{ab}=O(k^{M+1})\ ,\nonumber\end{eqnarray}from the quadratic nature of $\hat G^{ab}$. To obtain a solution of Einstein's field equations in the Nth. approximation, 
terminate the sequence $\{\underset{M}{\gamma_{ab}}\}$ at the $N$th. term by imposing the \emph{Integrability Conditions/Equations of Motion} in the $N$th. approximation: 
\begin{eqnarray}
\underset{N-1}{H}^{ab}{}_{,b}\equiv T^{ab}{}_{,b}+(8\,\pi)^{-1}\underset{N-1}{\hat G}^{ab}{}_{,b}=0\ .\nonumber\end{eqnarray}
Now
\begin{eqnarray}
\underset{N}{\gamma^*_{ab}}=-16\,\pi\,J\underset{N-1}{H}^{ab}\ ,\nonumber\end{eqnarray}and
\begin{eqnarray}
\underset{N}{G}^{ab}+8\,\pi\,T^{ab}=O(k^{N+1})\ ,\nonumber\end{eqnarray}
showing that Einstein's field equations are approximately satisfied in this sense. This scheme was subsequently utilised for the study of equations of motion in general relativity \cite{a}, \cite{b}, \cite{c} and \cite{d}.

In an amusing spin-off Synge \cite{Nature} constructed the following divergence--free pseudo--tensor: first write the vanishing covariant divergence of the energy--momentum--stress tensor in the equivalent forms
\begin{eqnarray}
T^{ab}{}_{|b}=0\ \ \Leftrightarrow\ \ T^{ab}{}_{,b}+K_a=0\ .\nonumber\end{eqnarray}
Here $K_a=\Gamma^a_{cb}\,T^{cb}+\Gamma ^b_{cb}\,T^{ac}$ is not a tensor (so the position of the index $a$ is not significant; $\Gamma^a_{bc}$ are the components of the Riemannian connection calculated with the 
metric tensor $g_{ab}$). Then define the pseudo--vector 
\begin{eqnarray}
Q_a=JK_a\ \ \Rightarrow\ \ \Box Q_a=K_a\ ,\nonumber\end{eqnarray} 
(with $J$ the operator introduced above and $\Box$ the Minkowskian d'Alembertian operator) and define the pseudo--tensor
\begin{eqnarray}
\varphi_{ab}=Q_{a,b}+Q_{b,a}-\delta_{ab}\,Q_{c,c}\ .\nonumber\end{eqnarray}It thus follows that
\begin{eqnarray}
\varphi_{ab,b}=\Box Q_a+Q_{b,ab}-Q_{c,ca}=\Box Q_a=K_a=-T^{ab}{}_{,b}\ .\nonumber\end{eqnarray}
Hence 
\begin{eqnarray}
\tau^{ab}=T^{ab}+\varphi_{ab}=\tau^{ba}\ ,\nonumber\end{eqnarray}
is \emph{a pseudo--tensor with vanishing divergence} ($\tau^{ab}{}_{,b}=0$). However Synge offered, in his characteristic style, these words of warning: ``I refrain from attaching the words momentum and energy to this pseudo--tensor or to integrals formed from it, because I believe that we are barking up the wrong tree if we attach such important physical terms to 
mathematical constructs which lack the essential invariance property fundamental in general relativity." 

\setcounter{equation}{0}
\section{Lorentz Transformations}\indent 
Synge gave a succinct description of his early education when he wrote \cite{Synge/AutobioNotes}:
``Although there are great gaps in my scientific equipment - like Hadamard, I could never get my teeth into group theory - I think I have ranged more widely than most. I might easily have stuck to classical subjects 
in which I was well trained as an undergraduate (dynamics, hydrodynamics, elasticity), but I wanted to take part in the new subjects, and in due course I mastered relativity but not quantum theory." True to this background, when considering Lorentz transformations, Synge thought of the analogy with ``the kinematics of a rigid body with a fixed point" (in \cite{RST}) and thus the construction of a general rotation in three dimensional Euclidean space in 
terms of the Euler angles. For Lorentz transformations the analogy requires six transformations of an orthonormal tetrad to another orthonormal tetrad, involving three pseudo angles (the arguments of hyperbolic functions) and three Euclidean angles. While this perspective is interesting the resulting formalism is not well suited to discussing the detailed effect of Lorentz transformations on the null cone. In the second edition of his text on special relativity Synge thanked I. Robinson and A. Taub ``for pointing out an error in Chapter IV of this book as first published (1955): singular Lorentz transformations were overlooked."

Taub was using spinors but Robinson had encountered the singular case in a novel way \cite{Ivor1,Ivor2,Ivor3}: Robinson was interested in the Schwarzschild solution in the limit $m\rightarrow+\infty$. Starting with the Eddington--Finkelstein form
\begin{eqnarray}
ds^2=-\frac{r^2\,(dx^2+dy^2)}{\left \{1+\frac{1}{4}(x^2+y^2)\right \}^2}+2\,du\,dr+\left (1-\frac{2\,m}{r}\right )du^2\nonumber\end{eqnarray}
and, using a clever coordinate transformation, Robinson wrote this in the form
\begin{eqnarray}
ds^2=-\frac{r^2}{\cosh^2\lambda\xi}(d\xi^2+d\eta^2)+2\,du\,dr+\left (\lambda^2-\frac{2}{r}\right )du^2\ ,\ \ \ \lambda=m^{-1/3}\nonumber\end{eqnarray}
Taking the limit $\lambda\rightarrow 0$ ($\Leftrightarrow m\rightarrow+\infty$) this becomes
\begin{eqnarray}
ds^2=-r^2(d\xi^2+d\eta^2)+2\,du\,dr-\frac{2}{r}du^2\nonumber\end{eqnarray}
This is another (different from Schwarzschild) Robinson--Trautman \cite{RT} type D vacuum space--time. The metric tensor has one term singular at $r=0$. This line element can be written in the form
\begin{eqnarray}
ds^2=-T^{4/3}(dX^2+dY^2)-T^{-2/3}dZ^2+dT^2\nonumber\end{eqnarray}
which is a Kasner \cite{K} solution of Einstein's vacuum field equations. If we remove the term singular at $r=0$ above we have a line element
\begin{eqnarray}
ds^2=-r^2(d\xi^2+d\eta^2)+2\,du\,dr\nonumber\end{eqnarray}
This is flat space--time and $r=0$ is a null geodesic. Hence
\begin{eqnarray}
\xi\rightarrow\xi+a\ ,\ \ \eta\rightarrow\eta+b\ ,\ \ u\rightarrow u\ ,\ \ r\rightarrow r\nonumber\end{eqnarray}
where $a, b$ are real constants, constitutes \emph{a Lorentz transformation leaving only the null direction $r=0$ invariant}. This is a singular Lorentz transformation (or null rotation) and the example moreover shows 
that such transformations exist and constitute a two--parameter Abelian subgroup of the Lorentz group. 

\setcounter{equation}{0}
\section{Synge on an Observation of E. T. Whittaker}\indent 
I mentioned at the outset that Professor Synge remained research active well into old age. To demonstrate this I want to give an example of some work carried out when he was eighty--eight years old. For several years, starting 
in the early 1980's, he and I found it convenient to correspond via letter. This allowed easy exchange of the results of calculations before the age of email. He typed his letters, including equations, on an ancient machine which he had used for years. The example I want to give involves an observation due to E. T. Whittaker and to do it justice I must first give a fairly extensive introduction.

Whittaker (in \cite{W1}, \cite{W2}) was concerned with the Li\'enard--Wiechert electromagnetic field of a moving charge $e$ so we will need some notation which we can briefly summarise as follows: 
\vskip 1truepc
\noindent
1) Line element: $ds^2=\eta_{ij}dX^i\,dX^j=-dX^2-dY^2-dZ^2+dT^2\ .$
\vskip 1truepc
\noindent
2) World line of charge: $X^i=w^i(u)\ ;\ v^i(u)=dw^i/du\ \ {\rm with}\ \ v^i\,v_i=+1$ ($\Rightarrow v^i=$ 4--velocity, $u=$ arc length or proper time); $a^i=dv^i/du=$ 4--acceleration $\ \Rightarrow\ a^i\,v_i=0$)
\vskip 1truepc
\noindent
3) Retarded distance of $X^i$ from $X^i=w^i(u)$: 
\begin{eqnarray}
r=\eta_{ij}(X^i-w^i(u))v^j\geq 0\ ;\ \eta_{ij}(X^i-w^i(u))(X^j-w^j(u))=0\ .\nonumber\end{eqnarray}
\vskip 1truepc
\noindent
Let $X^i-w^i(u)=r\,k^i$ then $k^i\,k_i=0$ and $k^i\,v_i=+1$. Parametrise the \emph{direction} of $k^i$ by $x, y$ such that
\begin{eqnarray}
k^i=P_0^{-1}\left (-x, -y, -1+\frac{1}{4}(x^2+y^2), 1+\frac{1}{4}(x^2+y^2)\right )\ ,\nonumber\end{eqnarray}
and then the normalisation $k^i\,v_i=+1$ implies
\begin{eqnarray}
P_0=x\,v^1(u)+y\,v^2(u)+\left\{1-\frac{1}{4}(x^2+y^2)\right\}v^3(u)+\left\{1+\frac{1}{4}(x^2+y^2)\right\}v^4(u)\ .\nonumber\end{eqnarray}

Whittaker observed that the Li\'enard--Wiechert 4--potential
\begin{eqnarray}
A^i=\frac{e\,v^i}{r}\ \ \Rightarrow\ \ A^i{}_{,i}=0=\Box A^i\ ,\nonumber\end{eqnarray}
could be written, modulo a gauge transformation, in the form
\begin{eqnarray}
A^i=\frac{e\,v^i}{r}=K^{ij}F_{,j}+{}^*K^{ij}G_{,j}\ ,\nonumber\end{eqnarray}
where $K^{ij}=-K^{ji}$ is a constant real bivector, with ${}^*K_{ij}=\frac{1}{2}\epsilon_{ijkl}\,K^{kl}$ its dual, and $F, G$ are real--valued functions each satisfying the Minkowskian wave equation
\begin{eqnarray}
\Box F=0 \ \ {\rm and}\ \ \Box G=0\ .\nonumber\end{eqnarray}

\noindent
To establish this in coordinates $x, y, r, u$ we need
\begin{eqnarray}
\frac{\partial}{\partial X^i}=-\frac{P_0^2}{r}\left (\frac{\partial k_i}{\partial x}\frac{\partial}{\partial x}+\frac{\partial k_i}{\partial y}\frac{\partial}{\partial y}\right )+v_i\frac{\partial}{\partial r}
+k_i\left\{\frac{\partial}{\partial u}-(1-r\,a_i\,k^i)\frac{\partial}{\partial r}\right\}\ ,\nonumber\end{eqnarray}
and
\begin{eqnarray}
\Box =-\frac{P_0^2}{r^2}\left (\frac{\partial^2}{\partial x^2}+\frac{\partial^2}{\partial y^2}\right )-(1-2\,a_i\,k^i\,r)\left (\frac{\partial^2}{\partial r^2}+\frac{2}{r}\frac{\partial}{\partial r}\right )+2\,\frac{\partial^2}{\partial u\partial r}+\frac{2}{r}
\frac{\partial}{\partial u}\ .\nonumber\end{eqnarray}
In coordinates $x, y, r, u$ Whittaker's two wave functions are \cite{EH}
\begin{eqnarray}
F=-\frac{e}{2}\log(x^2+y^2)\ \ \ {\rm and}\ \ \ G=-e\,\tan^{-1}\frac{y}{x}\ ,\nonumber\end{eqnarray}
(two \emph{harmonic} functions) and thus
\begin{eqnarray}
\frac{\partial F}{\partial X^i}=\frac{e\,P_0^2}{r\,(x^2+y^2)}\left (x\,\frac{\partial k_i}{\partial x}+y\,\frac{\partial k_i}{\partial y}\right )\ ,\nonumber\end{eqnarray}
and
\begin{eqnarray}
\frac{\partial G}{\partial X^i}=\frac{e\,P_0^2}{r\,(x^2+y^2)}\left (x\,\frac{\partial k_i}{\partial y}-y\,\frac{\partial k_i}{\partial x}\right )\ .\nonumber\end{eqnarray}
Define
\begin{eqnarray}
K^{ij}=\delta^i_3\,\delta^j_4-\delta^i_4\,\delta^j_3\ \ \ {\rm and}\ \ \ L^{ij}=\delta^i_1\,\delta^j_2-\delta^i_2\,\delta^j_1={}^*K^{ij}\ ,\nonumber\end{eqnarray}
then
\begin{eqnarray}
A^i=K^{ij}F_{,j}+{}^*K^{ij}G_{,j}=\frac{e\,v^i}{r}+\eta^{ij}\,\Psi_{,j}\ ,\nonumber\end{eqnarray}with
\begin{eqnarray}
 \Psi=e\,\log\{r\,P_0^{-1}\sqrt{x^2+y^2}\}\ .\nonumber\end{eqnarray}
Whittaker pointed out that this decomposition is analogous to the splitting of a plane light wave into two plane polarised components. A notable fact is that \emph{almost every vacuum Maxwell field can be resolved into two parts in this way}. The presentation of Whittaker's observation in coordinates $x, y, r, u$ facilitates the derivation of the explicit decomposition (see \cite{EH}) for the Goldberg--Kerr electromagnetic field \cite{GK}. The second, and final, part of 
the introduction, to enable us to appreciate Synge's contribution, involves a simple proof of this decomposition of a vacuum Maxwell field in general.

We are working in Minkowskian space--time and we shall write the line element as given above in rectangular Cartesian coordinates and time $X^i=(X, Y, Z, T)$ with $i=1, 2, 3, 4$. In addition we shall make use of the 
following basis vector fields:
 \begin{eqnarray}
k^i\frac{\partial}{\partial X^i}&=&\frac{\partial}{\partial Z}+\frac{\partial}{\partial T}\ ,\ l^i\frac{\partial}{\partial X^i}=-\frac{\partial}{\partial Z}+\frac{\partial}{\partial T}\ ,\nonumber\\
\ m^i\frac{\partial}{\partial X^i}&=&
\frac{\partial}{\partial X}+i\frac{\partial}{\partial Y}\ ,\ \bar m^i\frac{\partial}{\partial X^i}=\frac{\partial}{\partial X}-i\frac{\partial}{\partial Y}\ .\nonumber\end{eqnarray}
All scalar products (with respect to the Minkowskian metric) of the pairs of these vectors vanish except $k^i\,l_i=+2$ and $m^i\,\bar m_i=-2$. In what follows a complex self--dual bivector satisfies: $A_{ij}=-A_{ji}$ and ${}^*A_{ij}=iA_{ij}$ and a complex anti--self--dual bivector satisfies: $B_{ij}=-B_{ji}$ and ${}^*B_{ij}=-iB_{ij}$, with the star denoting the Hodge dual. A basis of complex \emph{anti--self--dual} bivectors is given by 
\begin{eqnarray}
m_{ij}=m_i\,k_j-m_j\,k_i\ ,\ n_{ij}=\bar m_i\,l_j-\bar m_j\,l_i\ ,\nonumber\end{eqnarray}
and
\begin{eqnarray}
l_{ij}=m_i\,\bar m_j-\bar m_i\,m_j+l_i\,k_j-l_j\,k_i\ .\nonumber\end{eqnarray}
Let $F_{ij}=-F_{ji}$ be a candidate for a real Maxwell bivector. Since $F_{ij}+i{}^*F_{ij}$ is an anti--self--dual complex bivector it can be expanded on the basis above as
\begin{eqnarray}
F_{ij}+i{}^*F_{ij}=\phi_0\,n_{ij}+\phi_1\,l_{ij}+\phi_2\,m_{ij}\ ,\nonumber\end{eqnarray}
where $\phi_0, \phi_1, \phi_2$ are complex--valued functions of $X^i$. Maxwell's Equations 
\begin{eqnarray}
(F^{ij}+i{}^*F^{ij})_{,j}=0\ ,\nonumber\end{eqnarray}
imply integrability conditions for the existence of a complex--valued function $Q(X^i)$ such that:
\vskip 1truepc\noindent
(a) $Q$ is a wave function: $\Box Q=0\ \ \Leftrightarrow\ \ \bar m^i\,m^j\,Q_{,ij}=k^i\,l^j\,Q_{,ij}$\ ;
\vskip 1truepc\noindent
(b) $\phi_0=\frac{1}{4}k^i\,m^j\,Q_{,ij}\ ,\ \phi_1=\frac{1}{4}k^i\,l^j\,Q_{,ij}\ ,\ \phi_2=-\frac{1}{4}l^i\,\bar m^j\,Q_{,ij}$\ .
\vskip 1truepc\noindent
Let $\bar l_{ij}$ denote the complex conjugate of $l_{ij}$, then $\bar l_{ij}$ is \emph{self--dual}. Define
\begin{eqnarray}
W_{ij}=\frac{1}{4}\bar l_i{}^p\,Q_{,pj}-\frac{1}{4}\bar l_j{}^p\,Q_{,pi}=-W_{ji}\ .\nonumber\end{eqnarray}Since $Q$ is a wave function it follows that $W_{ij}$ is \emph{anti--self--dual}. Expressing $W_{ij}$ on the 
anti--self--dual bivector basis, and using (b) above, results in 
\begin{eqnarray}
W_{ij}=F_{ij}+i{}^*F_{ij}\ .\nonumber\end{eqnarray} Hence with $\frac{1}{4}\bar l_{ij}=K_{ij}-i{}^*K_{ij}$ and $Q=U+iV$, we can write
\begin{eqnarray}
F_{ij}=A_{i,j}-A_{j,i}\ \ \ {\rm with}\ \ \ A_i=K_i{}^j\,U_{,j}+{}^*K_i{}^j\,V_{,j}\ .\nonumber\end{eqnarray}
Thus in general an analytic solution of Maxwell's vacuum field equations on Minkowskian space--time can be constructed from a pair of real wave functions $U, V$ and a constant real bivector $K_{ij}=-K_{ji}$. The 
classic paper on this type of result for zero rest mass, spin $s$ fields is that of Penrose \cite{Pen} (see also Stewart \cite{St}).

When I wrote out this proof (incorporated into \cite{PH1}) and sent it to Synge his response was characteristic. He worked it all out for himself and sent me the following proof in December, 1985:
 \vskip 1truepc\noindent
Synge's proof begins with  
\vskip 1truepc\noindent
{\it Lemma}: With $X^i=(X, Y, Z, T),  \eta_{ij}={\rm diag}(-1, -1, -1, +1),  F_{ij}=-F_{ji}$ a Maxwell field so that $F^{ij}{}_{,j}=0\ ; \ F_{j,k}+F_{ki,j}+F_{jk,i}=0$ then
\begin{eqnarray}
F_{ij}=0\ \ {\rm  at}\ \  T=0\ \ \Rightarrow\ \ F_{ij}=0\ \  {\rm for}\ {\rm all}\ \ T\ .\nonumber\end{eqnarray}
``You cannot make energy out of nothing" (Synge)
\vskip 1truepc\noindent
{\it Corollary}: If $F_{ij}$ and $H_{ij}$ are Maxwell fields then
\begin{eqnarray}
F_{ij}=H_{ij}\ \ {\rm at}\ \ T=0\ \ \Rightarrow\ \ F_{ij}=H_{ij}\ \ {\rm for}\ {\rm  all}\ \ T\ .\nonumber\end{eqnarray}
With these preliminaries Synge stated the following:
\vskip 1truepc
\noindent
{\it Theorem}: Given a Maxwell field $F_{ij}$ and 
\begin{eqnarray}
H_{ij}=K_i{}^l\,U_{,lj}+{}^*K_i{}^l\,V_{,lj}-K_j{}^l\,U_{,li}-{}^*K_j{}^l\,V_{,li}\ ,\nonumber\end{eqnarray}
with $K_{ij}=-K_{ji}={\rm constants}$ and $U, V$ wave functions, then $H_{ij}$ is a Maxwell field and there exists $K_{ij}, U, V$ such that 
\begin{eqnarray}
H_{ij}=F_{ij} \ \ {\rm at}\ \ T=0\ .\nonumber\end{eqnarray}
 \vskip 1truepc
\noindent
Comment: Clearly $H_{ij}$ is a solution of Maxwell's equations. The choice of $K_{ij}, U, V$ is not unique. The theorem demands only their existence.
\vskip 1truepc\noindent
{\it Proof}: Choose $K^{ij}=\delta^i_3\,\delta^j_4-\delta^i_3\,\delta^j_4$ then ${}^*K^{ij}=\delta^i_1\,\delta^j_2-\delta^i_2\,\delta^j_1$ and writing out $H_{ij}=F_{ij}$ at $T=0$ we find the following pairs of equations for the Cauchy data $U, V, U_{,4}, V_{,4}$ for the wave functions at $T=0$: (all equations evaluated at $T=0$)
\vskip 1truepc\noindent
(A):\ \ \ \ $(U_{,4})_{,1}=F_{13}+V_{,23}\ \ \ {\rm and}\ \ \ (U_{,4})_{,2}=F_{23}-V_{,13}$\ ;
\vskip 1truepc\noindent
(B):\ \ \ $ (V_{,4})_{,1}=F_{24}-U_{,23}\ \ \ {\rm and}\ \ \ (V_{,4})_{,2}=-F_{14}+U_{,13}$\ ;
\vskip 1truepc\noindent
(C):\ \ \ \  $U_{,11}+U_{,22}=-F_{34}\ \ \ {\rm and}\ \ \ V_{,11}+V_{,22}=-F_{12}$\ .
\vskip 1truepc\noindent
If the equations (A) are consistent and if the equations (B) are consistent then (A), (B) and (C) can in principle be solved for the Cauchy data. The consistency follows from the assumption that $F_{ij}$ is a Maxwell field since 
then (A) implies that 
\begin{eqnarray}
(U_{,4})_{,12}-(U_{,4})_{,21}=F_{13,2}-F_{23,1}+V_{,232}+V_{,131}=F_{13,2}+F_{32,1}+F_{21,3}=0\ ,\nonumber\end{eqnarray}
and (B) implies that
\begin{eqnarray}
(V_{,4})_{,12}-(V_{,4})_{,21}=F_{24,2}+F_{14,1}-U_{,232}-U_{,131}=F_{24,2}+F_{14,1}+F_{34,3}=0\ ,\nonumber\end{eqnarray}
and the theorem is established.

\setcounter{equation}{0}
\section{Epilogue}\indent 
When visitors came to the Center for Relativity in the University of Texas at Austin, Alfred Schild, the founder of the Center and one of Synge's former collaborators \cite{SS} would enthusiastically point out to them 
that this was where Roy Kerr found his solution. This raises the question: what were the stand--out works produced in Professor Synge's school of relativity in Dublin? I discussed this with George Ellis some time ago and 
we concluded that Felix Pirani's study of the physical significance of the Riemann tensor \cite{Pir} and Werner Israel's proof of the uniqueness of the static black hole (uncharged \cite{WI} and charged \cite{WI1}) are \emph{arguably} the most profound products of 
Synge's school.

When Synge turned ninety years of age a small conference was organised in his honour. His status within Ireland was reflected in the report in a national newspaper which stated: ``President Hillery [Head of State] attended a special event in the Dublin Institute for Advanced Studies yesterday to wish a happy 90th. birthday to Professor Emeritus J. L. Synge, Ireland's most distinguished 
mathematician of the present century. Although he has been retired for fifteen years, the professor, a nephew of the playwright J. M. Synge, published three papers last year and has two more 
at present in the course of publication" [\emph{Irish Times}, 23rd. March, 1987].

My photograph of Professor Synge (Figure 1 above) was taken in July, 1987 in my back garden. Also present were two of Synge's former students, Dermott Mc Crea (see \cite{a}, \cite{b} and \cite{d} for example)  and Stephen O'Brien (of the O'Brien--Synge junction conditions \cite{OBS}) together with  Bill Bonnor who was visiting from the University of London.

\bibliographystyle{unsrt}
\bibliography{hogan_relgeo_proceedings_2016}
\end{document}